\documentclass[aps,prb,twocolumn,amsfonts,amssymb,amsmath,floats,floatfix,showpacs,preprintnumbers,superscriptaddress, 10pt]{revtex4-1}
\usepackage{standalone} 

\usepackage{graphicx}
\usepackage{hyperref}
\usepackage{dcolumn}   
\usepackage{bm}        

\usepackage{ifthen}
\usepackage{pgfplots}
\usepgfplotslibrary{external}

\hypersetup{
colorlinks=true, 
urlcolor= blue, 
citecolor=blue,
linkcolor= blue, 
bookmarks=true, 
bookmarksopen=false, 
    pdfauthor   = {Giorgio Levy},
    pdftitle    = {Resolution ARPES},
    pdfkeywords = {ARPES, low-energy excitations}, }
\usepackage[pdftex]{thumbpdf}       
\newboolean{use_export_figures}
\setboolean{use_export_figures}{true}

\pgfplotsset{compat=1.3}
\pgfplotsset{tick style={thin,black}}

\newcommand{\figfolder}{./}

\newboolean{nofigures}
\setboolean{nofigures}{false}

\newcommand{\ifexptkzst}[1]{
\ifthenelse{\boolean{use_export_figures}}{\includegraphics{\figfolder #1}}{\input{\figfolder #1}}
}
\newcommand{\showfigures}[1]{
\ifthenelse{\boolean{nofigures}}{}{#1}}

\newcommand{\gsim}{\mathord{\sim}}

\def\app#1#2{%
  \mathrel{%
    \setbox0=\hbox{$#1\sim$}%
    \setbox2=\hbox{%
      \rlap{\hbox{$#1\propto$}}%
      \lower1.1\ht0\box0%
    }%
    \raise0.25\ht2\box2%
  }%
}

\definecolor{input}{RGB}{0, 0, 228}
\definecolor{fitted}{RGB}{222, 0, 52}
\definecolor{simul}{RGB}{26, 100, 65}
\newcommand{\glsimul}[1]{#1}
\newcommand{\glinput}[1]{#1}

\newcommand{\cvcomment}[1]{}
\newcommand{\adcomment}[1]{}
\newcommand{\figtitle}[1]{(Color online) #1}

\begin{document}

\title{Deconstruction of Resolution Effects in Angle-Resolved Photoemission}
\author{G.\, Levy}
\email{giorgio.levy@phas.ubc.ca}
\affiliation{Department of Physics {\rm {\&}} Astronomy, University of British Columbia, Vancouver, British Columbia V6T\,1Z1, Canada}
\affiliation{Quantum Matter Institute, University of British Columbia, Vancouver, British Columbia V6T\,1Z4, Canada}
\author{W.\, Nettke}
\author{B.\,M. Ludbrook}
\author{C.\,N. Veenstra}
\affiliation{Department of Physics {\rm {\&}} Astronomy, University of British Columbia, Vancouver, British Columbia V6T\,1Z1, Canada}
\author{A. Damascelli}
\affiliation{Department of Physics {\rm {\&}} Astronomy, University of British Columbia, Vancouver, British Columbia V6T\,1Z1, Canada}
\affiliation{Quantum Matter Institute, University of British Columbia, Vancouver, British Columbia V6T\,1Z4, Canada}

\date{\today}

\begin{abstract}
We study how the energy and momentum resolution of angle-resolved photoemission spectroscopy (ARPES) affects the linewidth, Fermi crossing, velocity, and curvature of the measured band structure. 
Based on the fact that the resolution smooths out the spectra, acting as a low-pass filter, we develop an iterative simulation scheme that compensates for resolution effects and allows the fundamental physical parameters to be accurately extracted. 
By simulating a parabolic band structure of Fermi-liquid quasiparticles,
we show that this method works for an energy resolution up to 100\,meV and a momentum resolution equal to twice the energy resolution scaled by the Fermi velocity. Our analysis acquires particular relevance in the hard and soft x-ray regimes, where a degraded resolution limits the accuracy of the extracted physical parameters, making it possible to study how the electronic excitations are modified when the ARPES probing depth increases beyond the surface.
\end{abstract}

\maketitle 

\section{Introduction}
 The electronic excitations at the surface of solids can differ from those in the bulk because the three-dimensional translational symmetry---inherent to the periodic arrangement of atoms that constitutes a solid---is broken.\cite{Matzdorf04082000,Echenique2004,Fournier:2010,Veenstra:2013}  This highlights the need for experimental techniques that can probe the evolution of the electronic excitations from surface to bulk, and provide reliable information about the bulk electronic structure. Angle-resolved photoemission spectroscopy (ARPES) can be such a probe, owing to the possibility of progressively increasing the probing depth by varying the photon energy from the UV to the soft and hard X-ray regimes. \citep{damascelli2004,fadley:2012} In addition to an increased bulk sensitivity, ARPES at high photon energies also enables the study of the fully-developed three dimensional dispersion in the bulk, extraction of element-specific electronic information by means of resonant photoemission spectroscopy, probing the quantum interference between the decay of photoexcited core-holes and the excitations around the Fermi level, and gaining access to free-electron final states for the photoexcitation process. \citep{fadley:2012} However, in varying the photon energy from the UV to X-ray regimes, and based on current technical capabilities of ARPES, we face a critical dichotomy for the experimental study of electronic excitations in novel complex materials: on the one hand, working with UV photons achieves the highest energy and momentum resolutions, but also provides the highest sensitivity to the surface electronic structure; on the other hand, the soft and hard X-ray regimes probe deeper into the bulk, avoiding potential surface-related complications,\citep{damascelli2004,fadley:2012} but with worse resolution.
  
In the UV-regime, the energy and angular resolutions $\Delta\omega\gsim 1$\,meV and $\Delta\theta\gsim 0.1^\circ$ achieved by ARPES allow the extraction of the electronic self-energy for electrons with binding energy $\omega\!<\!10$\,meV with respect to the Fermi energy $E_\text{F}$, \citep{PhysRevLett.105.046402, Tamai2013} and also the study of the opening of superconducting gaps as small as $\gsim 1$\,meV and their momentum dependence along the normal state Fermi surface. \citep{PhysRevLett.108.037002}  For example, the use of UV lasers has enabled the measurement of the superconducting gap of CeRu$_2$ with a record-high energy resolution of $\Delta\omega\!=\!0.36$\,meV.\citep{PhysRevLett.94.057001}  However, the information obtained in this regime is mainly representative of a material's surface due to the short inelastic mean free path of the photoexcited electrons.\citep{damascelli2004} Instead, soft and hard X-rays probe deeper into the bulk, but the resolution is degraded by a factor of 10-to-100 as compared to the UV-regime.  This resolution degradation affects the observed energy--momentum dispersion relation $\epsilon_k$ and electronic lifetime, and limits our ability to observe and analyse the low-energy ({\it i.e.} $\omega\!<\!0.2$\,eV) electronic excitations in solids. 

As for the origin of this resolution degradation, we note that in ARPES experiments the total energy resolution $\Delta\omega$ is given by the sum in quadrature of electron-analyzer and photon-beam contributions. In the soft and hard X-ray regime, which requires the use of synchrotron radiation to attain the necessary high photon flux and energy, the ultimate energy resolution is typically limited by the beamline monochromator contribution, $\Delta h\nu$, defined by its resolving power $R_m\!=\!h\nu/\Delta h\nu$. State-of-the-art soft X-ray beamlines can achieve a resolving power as good as $R_m\!\simeq\!33,000$ for photon energies $h\nu\!\simeq\!1$\,keV,  corresponding to an ultimate energy resolution of $\Delta h\nu\!\simeq\! 30$\,meV.\citep{Strocov2010}  As for the total momentum resolution $\Delta k$, this is mainly determined by the angular resolution of the detector and the kinetic energy of the photoelectrons.\citep{damascelli2004, momres} For a $\Delta\theta\!\simeq\!0.1^\circ$ angular resolution, the momentum resolution varies from $\Delta k\!\simeq\!4\!\times\! 10^{-4}\text{\AA}^{-1}$ in the UV-regime ($h\nu\!\simeq\!16$\,eV) to $3\!\times\! 10^{-3}\text{\AA}^{-1}$ in the X-ray regime ($h\nu\!\simeq\!900$\,eV).

Attempts to mitigate the effects of poor energy and momentum resolution on the determination of the underlying physical parameters of a system can be classified into three groups: 
 {i)} comparison between experimental results and theoretical calculations where the experimental resolutions are included; \citep{cui2010}
 { ii)} deconvolution methods, such as Lucy-Richardson or Wiener filters,\cite{lucy1} to reduce the effects of the resolution broadening
 before any further analysis is performed;\citep{lucy2,rameau2010}
 { iii)} the combination of a one-dimensional fitting routine with the convolution with an instrumental resolution function. \citep{ingle2005, PhysRevB.79.054517}
Methods in the first group involve a theoretical description of the excitations and are therefore model dependent, while those in the second require a high signal-to-noise ratio since otherwise they would be prevented altogether by the noise magnification during the deconvolution process. The third approach is based on a phenomenological description of the ARPES data; as compared to the other approaches it does not demand the development of a specific model and does not require as high a signal--to--noise ratio, and will therefore be the one followed here.

In this paper, we present a systematic study of how momentum and energy resolutions affect the observed dispersion and lifetime of the electronic excitations. By performing an analysis of momentum distribution curves (MDCs), obtained as constant energy cuts of the ARPES intensity data, we verify that the momentum resolution is responsible only for an energy-independent contribution to the MDC linewidth, provided it is smaller than the energy resolution scaled by the quasiparticle velocity (we also note that the MDC analysis is only valid for weakly momentum-dependent self-energies). This observation allows us to concentrate on the effects of the energy resolution alone: although the latter hampers a straightforward extraction of the physical quantities when it is larger than 25\,meV, we show that those can be recovered using an iterative algorithm, which belongs to the phenomenological third approach mentioned above. As will be discussed later, this new method -- called {\it iterative deconstruction algorithm (IDA)} -- is based on the observation that the main effect of the energy resolution is to act as a low-pass filter on the ARPES signal.

\section{Spectral function}

 We start by describing our phenomenological model. The intensity  $I(k, \omega)$ of the ARPES signal as a function of electron momentum $k$ and
 energy $\omega$ is written as: \citep{PhysRevB.67.064504, damascelli2004, ingle2005}
\begin{equation}
\resizebox{0.9\hsize}{!}{$I(k, \omega)= \left[ |M_{if}|^2 A(k, \omega) f(\omega, T)+B\right] \otimes R(\Delta k, \Delta \omega)$,} 
\label{eq:I}
\end{equation}
where $M_{if}$ represents  the matrix element which accounts for the selection rules for the optical transition between initial and final states, 
 $A(k, \omega)$ is the single-particle spectral function describing the electronic excitations in the solid, 
 $f(\omega, T)$ is the Fermi-Dirac distribution describing the statistical electronic population at temperature $T$ for states with energy $\omega$ with respect to the chemical potential, and $B$ is a background. These quantities are convolved with the instrumental resolution function $R(\Delta k, \Delta \omega)$, where $\Delta k$ and $\Delta \omega$ are the total energy and momentum experimental resolutions. In this study we will neglect  quantum interference effects\cite{Zhu2012, Zhu2014} due to the matrix elements $|M_{if}|^2$, which depend on photon polarization and energy, by assuming a constant value; this is equivalent to considering a system where only a single initial-to-final-state transition is allowed.
Similarly, we also assume a step-like background $B$ for simplicity. 

The spectral function $A(k, \omega)$, describing the single-particle excitation spectrum, can be written as:\citep{damascelli2004, Comin2013} 
 \begin{equation}
 	A(k, \omega) = \frac{1}{\pi}\frac{-\Sigma''(k,\omega)}{\left[\omega-\epsilon_k^b-\Sigma'(k, \omega)\right]^2+\left[ \Sigma''(k, \omega) \right]^2}, 
 \end{equation}
where the self-energy $\Sigma(k, \omega)=\Sigma'(k,\omega)+i\Sigma''(k,\omega)$ captures the many-body correlation effects on the electronic excitations, and $\epsilon_k^b$ represents the bare-band dispersion. The effects of the self-energy are two-fold:\citep{damascelli2004, Comin2013}  the real part of the self-energy renormalizes the bare-band dispersion $\epsilon_k^b$ into the quasiparticle dispersion $\epsilon_k^q=\epsilon_k^b-\Sigma'(k, \omega)$, and the imaginary part $\Sigma''(k, \omega)$ describes the reduction in the lifetime of the single-particle excitations and the corresponding increase of the peak width in energy. 
For a weakly momentum-dependent self-energy, the spectral function may be further simplified by replacing $\Sigma(k, \omega)$ with $\Sigma(\omega)$, thus obtaining:
\begin{equation}
	A_\omega(k) = \frac{A_0}{\pi}
        \frac{\Delta k_m}
        {
          \left[ k-k_m(\epsilon) \right]^2
         +
          \left[ \Delta k_m \right]^2
        },
	\label{eq:SF}
\end{equation}
where $\Delta k_m$ is the half width at half maximum of a Lorentzian of weight $A_0$ centered at $k_m$ and determined from the apparent quasiparticle dispersion $\epsilon_k^q$.
In this case the electron self-energy may be extracted more straightforwardly from the ARPES spectra through Lorentzian fits of the MDCs, \citep{Valla,damascelli2004} even without any {\it a priori} knowledge of the bare-band $\epsilon_k^b$.\citep{PhysRevB.82.012504, PhysRevB.84.085126} However, the apparent quasiparticle dispersion $\epsilon_k^q$ and peak widths determined by $\Delta k_m=-\Sigma''(\omega)/ v_k^b$, where $v_k^b=\partial\epsilon_k^b/\partial k$ is the bare-band velocity, will be affected by both momentum and energy experimental resolutions. Note that in the following, we characterize the momentum linewidth by the full width at half maximum (FWHM) $\Gamma=2\Delta k_m$.

We establish here an analogy between the instrumental resolution $R(\Delta k, \Delta \omega)$ and a low-pass filter by considering the influence of resolution in the detection process. When an electron with  energy $\omega$ and momentum $k$ enters the detector, the instrumental resolution is determined by the probability of detecting it with energy $\omega'$  and momentum $k'$.  This probability distribution, represented by $R(\Delta k, \Delta \omega)$ in Eq. \ref{eq:I}, decays as $|\omega\!-\!\omega'|$ and $|k\!-\!k'|$ increase. As a result, the ARPES signal is proportional to the photoemitted electron distribution convolved with the instrumental resolution function $R(\Delta k, \Delta \omega)$. The effect of the instrumental resolution can be modeled as a {\it low-pass filter} because the resolution effectively smooths out the spectra, suppressing variations of the signal that have a frequency in energy/momentum higher than the resolution itself. Furthermore, the functional form of the experimental resolution $R(\Delta k, \Delta \omega)$ can be approximated by a Gaussian profile,\citep{imhof1976, storer1994} and within this approximation the instrumental resolution acting upon the signal is equivalent to a Gaussian filter.

One can show that, when the broadening due to the energy resolution scaled by the quasiparticle velocity is larger than the corresponding broadening due to the momentum resolution, the main effect of the momentum resolution in the MDC analysis is to increase the effective linewidth by an energy-independent value (Section\,\ref{Appendix:kres}). In particular, for momentum-independent self-energies the MDC lineshape can be described by a Lorentzian profile, \citep{PhysRevB.69.144509} which is modified into a Voigt profile by the convolution with a Gaussian resolution function in momentum. \citep{Hulst1947} Both curves are difficult to distinguish experimentally for low-to-medium signal-to-noise ratios since the largest difference is in the tails of the profiles, away from the peak position.
For this reason, we first restrict the analysis to the case $\Delta k=0$; we will discuss the effects of a finite momentum resolution later (Section\,\ref{Appendix:kres}) when we describe the IDA approach. Note, 
 however, that we have verified that a finite $\Delta k$ does not alter the results 
 when $\Delta k\leq 2\Delta \omega/v^q_\text{F}$, where $v_\text{F}^q$ is the quasiparticle Fermi velocity. We also note that this upper limit on the momentum resolution does not imply an experimental limitation on the maximum value of $v_\text{F}^q$ that can be measured by this technique because, for a given momentum resolution, higher values of $v_\text{F}^q$ can be accessed by decreasing (worsening) experimentally the energy resolution. Also, this condition does not imply that different results should be obtained between UV and X-ray regimes because energy and momentum resolutions scale approximately at the same rate with photon energy and, in turn, the resolution ratio $\Delta \omega/\Delta k$ remains comparable.

\begin{figure}[htb]
\showfigures{\input{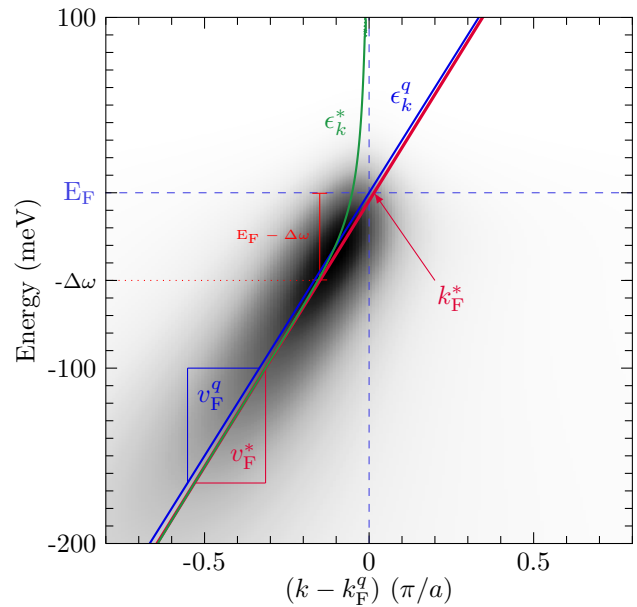}}
\caption{\label{fig1} \figtitle{Effect of energy resolution on a linear dispersion with Fermi-liquid linewidth.} The gray scale ARPES intensity is obtained from Eqs.\,\ref{eq:I} and \ref{eq:SF} for $T\!=\!10$\,K, $\Delta \omega\!=\!50$\,meV, and $\Delta k\!=\!0$, using the quasiparticle dispersion $\epsilon_k^q= v^q_\text{F} (k-k^q_\text{F})$ (blue line) and linewidth $\Gamma(\omega)=\Gamma_0+\Gamma_2\,\omega^2$, with $v^q_\text{F}\!=\!1\,\textnormal{eV}\,a/\pi$, $\Gamma_0\!=\!0.1\,\pi/a$, and $\Gamma_2\!=\!10\,\textnormal{eV}^{-2}\,\pi/a$ (all quantities including $k$ expressed in units of $\pi/a$). The extracted dispersion $\epsilon^*_k$ (green line) is obtained from the Lorentzian fit of the MDCs; Fermi  momentum $k^*_\text{F}$ and velocity $v_\text{F}^*$ from a linear fit of $\epsilon^*_k$ up to $E_\text{F}-\Delta \omega$ (red line).}
\end{figure}

Under these conditions -- and contrary to the case of momentum resolution -- the energy resolution modifies the energy dependence of the parameters obtained from the MDC analysis since it mixes spectral weight from states at different energies. We find that the effects of the energy resolution $\Delta \omega$, as depicted in Fig.\,\ref{fig1}, are:
{i)} a distortion of the functional form of the resulting dispersion $\epsilon_k^*$ with respect to the intrinsic $\epsilon_k^q$ in an energy region $\Delta \omega$ around $E_\text{F}$, as was reported previously;\citep{ingle2005}
{ii)} a shift of the peak positions close to $E_\text{F}$, which results in a different Fermi crossing $k_\text{F}^*$;\citep{PhysRevLett.83.5551, PhysRevB.64.094513}
{iii)} a modification of the Fermi velocity from $v_\text{F}^q$ to $v_\text{F}^*$. 
Additional effects not depicted in Fig.\,\ref{fig1} are an increase of MDC linewidth, $\Gamma(\omega)$, inversely proportional to the slope of the band dispersion, and a reduction of the average-rate-of-change of the linewidth with energy. This will lead to an energy-dependent increase of the MDC linewidth, which is in turn responsible for the fact that EDC and MDC analyses of ARPES spectra return different quasiparticle dispersions. \citep{ingle2005}

\section{Resolution effects \label{sec:resolution}}

\subsection{Linear quasiparticle dispersion}

To illustrate the resolution effects alluded to above, we simulate the spectral function $A(k, \omega)$ using a linear quasiparticle band $\epsilon_k^q= v^q_\text{F} (k-k^q_\text{F})$ with a corresponding Fermi-liquid energy-dependent width $\Gamma(\omega)=\Gamma_0+\Gamma_2\,\omega^2$, where $\Gamma_0$ and $\Gamma_2$ account for impurity and electron-electron scattering, respectively. As described in Eq.\,\ref{eq:I}, the spectral function is multiplied by the Fermi-Dirac distribution and then convolved with a Gaussian energy resolution function with unit-area and full-width half-maximum (FWHM) equal to $\Delta \omega$. An example of such simulation is shown as a gray scale plot in Fig.\,\ref{fig1}, with the linear dispersion $\epsilon_k^q$ in blue. Here we used an energy resolution $\Delta \omega\!=\!50$\,meV, a momentum resolution $\Delta k \!=\!0$, a Fermi velocity $v^q_\text{F}\!=\!1\,\textnormal{eV}\,a/\pi$, and linewidth parameters $\Gamma_0\!=\!0.1\,\pi/a$ and $\Gamma_2\!=\!10\,\textnormal{eV}^{-2}\,\pi/a$. We note that throughout the paper all dispersion-related quantities -- including the electron momentum $k$ -- are expressed in units of $\pi/a$, where $a$ is the lattice parameter; also, we set $T=10\,K$ for all simulations. 

\begin{figure}[!t]
\showfigures{\input{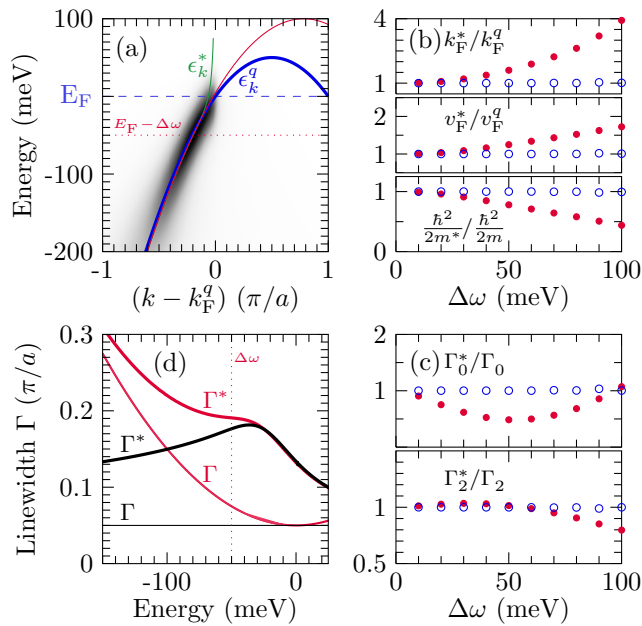}}
\caption{\label{figqband} \figtitle{Effect of energy resolution on a parabolic dispersion with Fermi-liquid linewidth.} (a) The gray scale ARPES intensity is obtained from Eqs.\,\ref{eq:I} and \ref{eq:SF}, for $T\!=\!10$\,K, $\Delta \omega\!=\!50$\,meV, and $\Delta k\!=\!0$, using the quasiparticle dispersion $\epsilon_k^q\!=\!\frac{\hbar^2}{2m}(k-k^q_\text{F})^2+v^q_\text{F}(k-k^q_\text{F})$ (blue line) and linewidth $\Gamma\!=\!\Gamma_0\!+\!\Gamma_2\omega^2$, with $v^q_\text{F}\!=\!0.2\,\text{eV}\,a/\pi$, $\Gamma_0\!=\!0.1\,\pi/a$, $\hbar^2/2m\!=\!-0.2\,\text{eV}\,(a/\pi)^2$, $\Gamma_0\!=\!0.1\,\pi/a$, and $\Gamma_2\!=\!10\,\text{eV}^{-2}\,\pi/a$ (all quantities including $k$ expressed in units of $\pi/a$). 
The green line is the dispersion $\epsilon_k^*$ extracted from the MDC analysis; the red one is the result of a quadratic fit of $\epsilon_k^*$ at binding energies deeper than $E_\text{F}-\Delta \omega$.
 (b) Energy-resolution dependence of the parameters (solid red symbols) $k_\text{F}^*$, $v_\text{F}^*$, and $\hbar^2/2m^*$, and (c) $\Gamma_0^*$ and $\Gamma_2^*$, as determined by fitting the band dispersion $\epsilon_k^*$ [green line in (a)] obtained from an MDC analysis of the ARPES intensity in (a); open symbols are the results of the IDA method discussed in Sec.\,\ref{sec:iterative}. 
The axis title is included inside the panels of (b) and (c).
 (d) The linewidth $\Gamma^*$ deviates from the input linewidth $\Gamma$ (red lines), even in the case of a purely constant input $\Gamma\!=\!\Gamma_0$ (black lines). 
\label{fig:qbandqwidth}}
\end{figure} 

Fitting the corresponding MDCs with a Lorentzian profile we obtain the dispersion \glsimul{$\epsilon_k^*$} (green line in Fig.\,\ref{fig1}), which is identical to the quasiparticle dispersion $\epsilon_k^q$ only when $\Delta \omega\!=\!0$. When the energy broadening part of the resolution function is larger than the width of the Fermi-Dirac distribution ($\Delta \omega\!>\!4k_B T$),\citep{ingle2005} $\epsilon_k^*$ deviates from the linearly dispersive band $\epsilon^q_k$ for energies closer to $E_\text{F}$ than $\Delta \omega$, showing an upturn above this energy. The interplay of energy resolution and spectral cut-off due to the Fermi-Dirac distribution is at the origin of these effects, as well as of the spectral weight induced above $E_\text{F}$ and extending up to $E_\text{F}+\Delta \omega$. Practically, the deviation of $\epsilon_k^*$ from $\epsilon_k^q$ in the range $|E_\text{F}- \Delta \omega|$ defines the maximum binding energy at which the band dispersion can be accurately traced; this also provides a method to estimate the energy resolution directly from the data: $\Delta \omega$ corresponds to the energy relative to $E_\text{F}$ where the upturn in $\epsilon_k^*$ has its onset.
 
At binding energies below the $\Delta \omega$ range around $E_\text{F}$, the extracted band dispersion \glsimul{$\epsilon^*_k$} is linear but is shifted compared to \glinput{$\epsilon_k^q$}, as shown by the comparison of green and blue lines in Fig.\,\ref{fig1}.  This shift is caused by the interplay of the energy resolution with the quadratic energy-dependence of the MDC linewidth,  which induces an asymmetry in the MDC profiles; when this asymmetric lineshape is fitted with a Lorentzian function, the result is a shift in peak position. This would not occur if there were only an energy-independent term in the momentum width $\Gamma\!=\!\Gamma_0$, or if the energy resolution broadening were reduced to $\Delta \omega\!=\!0$ (as shown below, a similar shift of the peak positions also occurs for non-linear dispersions). As reported previously,\citep{PhysRevLett.83.5551, PhysRevB.64.094513} in Fig.\,\ref{fig1} we can also see that as a consequence of this shift, the extrapolated Fermi momentum $k_\text{F}^*$ moves with respect to $k_\text{F}^q$, affecting the determination of the Fermi surface in an MDC analysis [this will be discussed in greater detail in relation to Fig.\,\ref{fig:qbandqwidth}(b)].

Next we study the variation, due to the energy resolution, of the extracted Fermi velocity $v_\text{F}^*$, as obtained from a linear fit of the dispersion $\epsilon_k^*$ up to $E_\text{F}-\Delta \omega$. The deviation of the Fermi velocity $v_\text{F}^*$ from the intrinsic $v^q_\text{F}$ depends on the interplay of temperature $T$, linewidth $\Gamma$, and energy resolution $\Delta \omega$. Note that for zero energy resolution, $v_\text{F}^*\!=\!v^q_\text{F}$ independent of the other parameters, which demonstrates that its deviation is due to a finite $\Delta \omega$. The relative velocity $v_\text{F}^r=v_\text{F}^*/v^q_\text{F}$ increases quadratically with the energy resolution; and for a given energy resolution, $v_\text{F}^r$ increases quadratically with temperature, semi-logarithmically with the input Fermi velocity $v^q_\text{F}$, and semi-logarithmically with the energy-independent momentum width $\Gamma_0$. In absolute terms, the deviations of $v_\text{F}^*$ due to temperature (up to 100\,K) and  energy-independent momentum width term $\Gamma_0$ (up to $2\Delta \omega/v^q_\text{F}$) are at most $6\%$. As expected, the largest contribution is due to $\Delta \omega$. 
The increase of the extracted Fermi velocity $v_\text{F}^*$ with energy resolution $\Delta \omega$ can be understood by the smoothing effect mentioned in the low-pass filter analogy: $\Delta \omega$ introduces an effective cut-off for the maximum rate-of-change observable in the energy distribution curves (EDCs); this EDC broadening, together with the quadratic energy dependence of the linewidth, translates into an increase of $v_\text{F}^*$ inferred from the MDC analysis. As a limiting case, we expect that $v_\text{F}^* \rightarrow \infty$ when $\Delta \omega \rightarrow \infty$. 

\subsection{Quadratic quasiparticle dispersion \label{sec:quad}}

Following the same approach, we also expect that the effect of energy resolution on a parabolic band dispersion is to reduce its curvature. This point is exemplified in Fig.\,\ref{fig:qbandqwidth}, where we consider the change of the extracted spectroscopic quantities due to energy resolution for the quasiparticle dispersion $\epsilon_k^q\!=\!\frac{\hbar^2}{2m}(k-k^q_\text{F})^2+v^q_\text{F}(k-k^q_\text{F})$ and a Fermi-liquid linewidth $\Gamma\!=\!\Gamma_0\!+\!\Gamma_2\omega^2$. From the ARPES intensity $I(k, \omega)$ in Fig.\,\ref{fig:qbandqwidth}(a), calculated for the parameter values indicated in the caption, we extract the quasiparticle dispersion $\epsilon_k^*$ (green line) affected by the energy resolution; this can be tracked up to $E_\text{F}-\Delta \omega$ before it deviates from the intrinsic $\epsilon_k^q$ dispersion.
By fitting $\epsilon_k^*$ with a parabolic dispersion outside of the $\Delta \omega$ energy range, we obtain the red line in Fig. \ref{fig:qbandqwidth}(a), from which we can extract estimates for the
 Fermi momentum $k_\text{F}^*$, Fermi velocity $v_\text{F}^*$, and quasiparticle curvature $\hbar^2/2m^*$. As shown by the red filled symbols in Fig.\,\ref{fig:qbandqwidth}(b), these extracted parameters vary quadratically with energy resolution relative to the input values $k^q_\text{F}$, $v^q_\text{F}$, and $\hbar^2/2m^*$, which are instead recovered for $\Delta \omega\!=\!0$. The variation of the extracted Fermi velocity $v_\text{F}^*$ and curvature $\hbar^2/2m^*$ with $\Delta \omega$ follows the guideline previously stated that the energy resolution tends to smooth out the ARPES spectra: it reduces the overall curvature and increases the Fermi velocity [see Fig.\,\ref{fig:qbandqwidth}(b)]. As a result,
the extracted Fermi momentum $k_\text{F}^*$ (decreases) increases with $\Delta \omega$ for (electron-) hole-like Fermi surfaces.

As for the energy-resolution dependence of the extracted linewidth $\Gamma^*\!=\!\Gamma_0^*+\Gamma^*_2\omega ^2$, in Fig.\,\ref{fig:qbandqwidth}(c) we observe that the relative variation of the energy-independent $\Gamma_0^*$ and quadratic $\Gamma_2^*$ are non-monotonic with the energy resolution $\Delta \omega$ (see red filled symbols, and again $\Delta \omega\!=\!0$ for the input values). In addition, the energy-dependence of $\Gamma^*$ is modified by a term inversely proportional to the slope of the quasiparticle dispersion $\epsilon_k^q$. To illustrate this, we consider first an energy-independent momentum width $\Gamma\equiv\Gamma_0$ as a simpler case. The MDC Lorentzian profile of width $\Gamma_0$ is modified  into a Voigt lineshape by the convolution with a Gaussian function of width $\sigma_g(E)\propto\Delta \omega / (\partial\epsilon_k/\partial k)|_{\omega}$. Remarkably [see black lines in Fig.\,\ref{fig:qbandqwidth}(d)], the extracted $\Gamma^*$ exhibits an energy dependence although the input linewidth $\Gamma=\Gamma_0$ is constant in energy. In case of the quadratic linewidth $\Gamma\!=\!\Gamma_0+\Gamma_2\omega ^2$ [see red lines in Fig.\,\ref{fig:qbandqwidth}(d)], the extracted $\Gamma^*$ still presents a parabolic dependence on energy for $\omega<E_\text{F}-\Delta\omega$, but deviates considerably from the input linewidth $\Gamma$. 

\section{Iterative deconstruction algorithm (IDA)}
\label{sec:iterative}

\begin{figure}[htb]
\showfigures{\includegraphics[width=8.75cm]{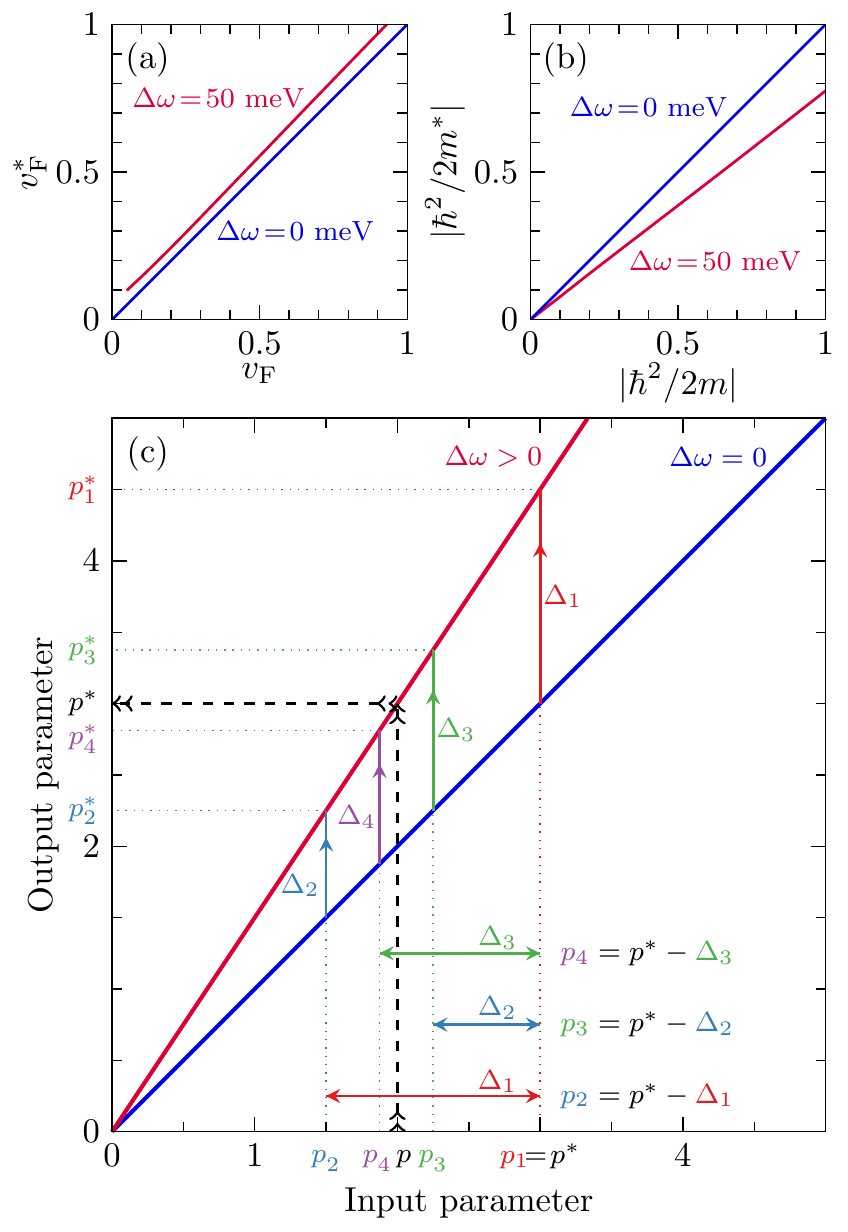}}
\caption{
\label{fig:iter} 
\figtitle{Iterative deconstruction algorithm (IDA) to retrieve the intrinsic parameters $p$ from the $p^*$ as {\it measured} -- and modified by the energy resolution.} At a constant energy resolution $\Delta \omega\!=\!50 $\,meV, the energy-resolution-induced deviations of (a) Fermi velocity $v_\text{F}^*$ and (b) band curvature $|\hbar^2/2m^*|$ from the corresponding input values increase with the magnitude of the latter; note that here all other parameters are the same as in Fig.\,\ref{fig:qbandqwidth}(a). The IDA method is illustrated in (c), and shows that starting from the {\it measured} parameters $p_1 \equiv p^*$, the iterative input ($p_i$) and extracted ($p_i^*$) parameters progressively converge towards the true ($p$) and measured values ($p^*$); note that here $\Delta_i=p_i-p_i^*$. The iteration is stopped when the difference between $p^*_i$ and $p^*$ is smaller than an appropriate tolerance factor.}
\end{figure}

So far, we have analysed the variation with energy resolution of the quantities $k_\text{F}^*$, $v_\text{F}^*$, $\hbar^2/2m^*$, $\Gamma_0^*$, and $\Gamma_2^*$, which parametrize the electronic dispersion $\epsilon_k^*$ and lifetime $\Gamma^*$ obtained from an MDC analysis of the ARPES intensity. We have shown that these variations can be understood as a cut-off on the maximum rate-of-change of the ARPES intensity imposed by the experimental resolutions, in analogy with a low-pass filter effect. This becomes more pronounced the stronger the energy dependence of the quasiparticle dispersion. For example, as shown in Fig.\,\ref{fig:iter}(a) and (b), the resolution-induced deviations of the extracted Fermi velocity $v_\text{F}^*$ and especially curvature $|\hbar^2/2m^*|$ increase with the input parameters; this observation can be generalized to each dispersion parameter $p$, to show that the absolute difference between intrinsic and extracted values, $|p\!-\!p^*|$, increases with $p$.  

Based on this observation we devise an iterative method to retrieve the intrinsic parameters $p$, starting from the $p^*$ extracted -- and affected by the experimental resolution -- through the MDC analysis of the {\it measured} ARPES intensity $I(k,\omega)$. The iterations are initialized defining the first set of parameters to be identical to the {\it measured} ones, $p_1 \equiv p^*$. Next, using this set of $p_1$ and the known energy resolution, a {\it simulated} ARPES intensity map is generated; new values for the parameters $p_1^*$ can then be extracted, now through an MDC analysis of the {\it simulated} intensity. Note that these newly determined $p_1^*$ will be further away from the intrinsic parameters $p$ than the input values $p_1 \equiv p^*$, due to the energy resolution broadening having effectively been accounted for twice. By taking the difference $\Delta_1\!=\!p_1-p_1^*$, and subtracting it from the {\it measured} $p^*$, we define the starting parameters for the next iteration: $p_{2}\!=\!p^*-\Delta_1$. As a result of the second iteration, we obtain the new difference $\Delta_2\!=\!p_2-p_2^*$ and then the input values for the third iteration: $p_3\!=\!p^*-\Delta_2$. These iterations are repeated until the difference between the output values $p_i^*$ and the {\it measured} $p^*$ is below a chosen tolerance factor. At this point, the input parameters $p_i$ of the last iteration, can be considered representative of the true values $p$. The key  iterative steps, with $i\!=\!1, 2, ...$, are thus
\begin{eqnarray}
p_1&=&p^*, \nonumber \\
\Delta_i &=& p_i - p_i^*, \\
p_{i+1} &=& p^* - \Delta_{i}. \nonumber
\end{eqnarray}

Note that the difference $\Delta_i$ is always combined with the {\it measured} parameters $p^*$ for all iterations; this is necessary because the information of the true values $p$ is encoded in the measured values $p^*$ together with the energy resolution $\Delta \omega$. This IDA method is illustrated in Fig.\,\ref{fig:iter}(c), and shows how the input ($p_i$) and extracted ($p_i^*$) parameters progressively converge towards the true ($p$) and measured ($p^*$) values. By applying the IDA to the example discussed before of a parabolic band dispersion with a Fermi-liquid linewidth, we find that the systematic error induced by the energy resolution is reduced to $<3\%$, as shown by the open circles in Fig.\,\ref{fig:qbandqwidth}(b) and (c). 
Finally, one should note that this method only relies on the monotonous increase of $|p_i-p_i^*|$ with $p_i$, and not on its specific functional form. 

\begin{figure}[!t]
\showfigures{\includegraphics{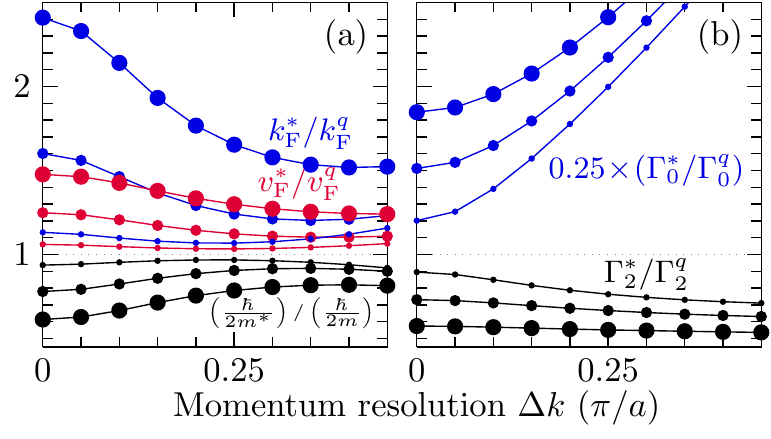}}
\caption{
\label{fig:kresf} 
\figtitle{Relative deviation of the MDC-extracted (a) quasiparticle dispersion and (b) linewidth parameters, plotted as a function of momentum resolution and for 25, 50, and 75\,meV energy resolution values (as represented by the size of the symbols). The input values for the simulations are the same as in Fig.\,\ref{fig:qbandqwidth}, and the extracted parameters are determined by the procedure described in Section\,\ref{sec:quad}. Note that the evolution of the energy-independent linewidth $\Gamma_0$ has been rescaled down by 0.25 in the plot.}
}
\end{figure}

It is important to note that the IDA systematic approach assures an improved rate of convergence -- for the same accuracy on the solution -- as compared to other methods, such as least-squares fitting, maximum likelihood estimation, or direct evaluation over a mesh of trial values for $p$. There are two reasons for this. The first one is the choice of the IDA starting parameters $p$, which are taken to be identical to the experimentally determined ones $p^*$. In contrast, the other methods require an initial guess of the parameters for the first iteration, and the rate of convergence to the true physical parameters $p$ is critically dependent on this initial guess. The second reason is that the IDA selects a specific path in search space, i.e. the space formed by all the possible values of $p$, reducing the number of evaluations per iteration to a single one. Conversely, the other methods explore the search space to find an optimal solution at each given iterative step, and therefore need more than one evaluation per iteration. As an illustration of the efficiency of the IDA approach, we note that for the cases studied here convergence was achieved in less than ten iterations (with only one evaluation per iteration).
\vspace{-2mm}
\section{Momentum resolution \label{Appendix:kres}}

\begin{figure}[tb]
\showfigures{\includegraphics{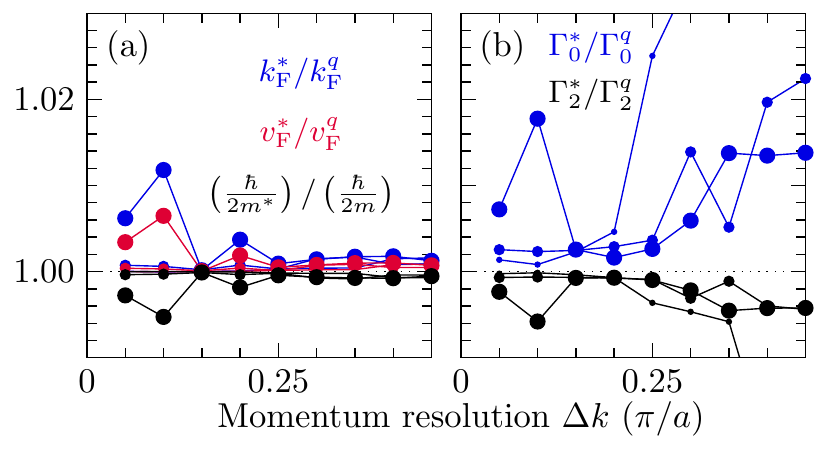}}
\caption{
\label{fig:kres} 
\figtitle{When the momentum resolution is included in the IDA method, the relative deviation on the extracted (a) quasiparticle dispersion and (b) linewidth pa\-ra\-me\-ters as a function of momentum resolution (for energy resolutions $\Delta \omega =$ 25, 50, and 75\,meV, as represented by the symbol size) is reduced in comparison to the results presented in Fig.\,\ref{fig:kresf}.}
}
\end{figure}

We have shown that the IDA method can compensate for energy resolution effects inasmuch as we neglect the momentum resolution. But since the instrumental resolution is compounded by both energy and momentum contributions, we need to include the latter in our analysis. 
Note that here we focus on the instrumental momentum resolution $\Delta k$ as defined in Eq.\,\ref{eq:I}, and we neglect the intrinsic momentum resolution due to the finite mean free path of the photoemitted electrons.\citep{Strocov200365}
Fig.\,\ref{fig:kresf} shows the modifications of the MDC-extracted Fermi crossing $k_\text{F}^*$, Fermi velocity $v_\text{F}^*$, quasiparticle curvature $\hbar/2m^*$, and Fermi-liquid linewidth $\Gamma^*=\Gamma_0^*+\Gamma_2^2\omega$ caused by a finite momentum resolution $\Delta k>0$, for energy resolution values of 25, 50 and 75\,meV.
 As $\Delta k$ increases from zero, the relative deviation of the extracted quasiparticle parameters-- Fermi crossing, Fermi velocity, and curvature-- actually approaches unity [Fig.\,\ref{fig:kresf}(a)]. 
 This may seem counter-intuitive, but is the result of the net effective compensation of the effects from the energy and momentum resolution. For instance, in the case of the extracted Fermi velocity $v_\text{F}^*$, the energy resolution will reduce it but the momentum resolution will increase it. 
 Similar behaviour is observed for the other parameters, with the exception of the  energy-independent linewidth parameter $\Gamma_0$ which increases with $\Delta k$ independent of $\Delta \omega$.

 For the parameters utilized in Fig.\,\ref{fig:kresf}, the relative deviation on the extracted parameters can be substantially reduced when a finite momentum resolution is accounted for in the IDA method, from $\gsim 50\%$ the MDC analysis (Fig.\,\ref{fig:kresf}) to even $< 3\%$ (Fig.\,\ref{fig:kres}). In fact deviations in excess of $3\%$ are observed only for the linewidth when $\Delta \omega \!=\!25$ meV and $\Delta k\!>\!0.25~(\pi/a)$. 
  Based on this, we can conclude that the extracted linewidth parameters are reliable for  momentum resolutions $\Delta k \leq 2\Delta \omega/v_\text{F}$. 
 When this condition applies, the IDA method can compensate for the deviation of dispersion $\epsilon_k^*$ and linewidth $\Gamma^*$ due to the combined inclusion of momentum ($\Delta k$) and energy ($\Delta \omega$) resolutions.

\vspace{-2mm}
\section{Conclusions}

We have systematically studied the effect of energy resolution on the measured MDC linewidth and quasiparticle dispersion parameters, such as Fermi velocity, Fermi momentum, and band curvature, as extracted from ARPES data. In particular, we considered the case of linear and parabolic dispersions, with a quadratic Fermi-liquid-like scattering rate. Starting from the observation that the energy resolution acts as a low-pass filter, we developed an iterative deconstruction algorithm to extract the underlying physical parameters, compensating for the progressive loss of energy resolution upon increasing of photon energy from the UV to hard X-ray regime. Based on these functional forms for the dispersion and scattering rate, this method provides an avenue for studying the electronic excitations with enhanced bulk sensitivity and to follow their bulk-to-surface evolution, with the highest degree of fidelity arbitrarily close to $E_\text{F}$ -- even closer than the energy resolution itself. Note however that this method relies on the trend of the MDC lineshape at energies at least twice as large as the energy resolution $\Delta \omega$; therefore, it cannot provide information on features confined to an energy scale smaller than $\Delta \omega$. This method can be generalized to other parametrizations and techniques where energy resolution produces similar effects, e.g. angle-resolved bremsstrahlung isochromat spectroscopy; however, its applicability should be verified case by case. 
\vspace{-2mm}
\begin{acknowledgements}

We thank G.A. Sawatzky and I.S. Elfimov for their critical questions which motivated this study.
This work was supported by the Killam, Alfred P. Sloan, and NSERC's Steacie Memorial Fellowships (A.D.), the Canada Research Chairs Program (A.D.), NSERC, CFI, and CIFAR Quantum Materials. 

\end{acknowledgements}

\appendix

\section{Measurement error in IDA \label{Appendix:Noise}}

\begin{figure}[!t]
\showfigures{\includegraphics{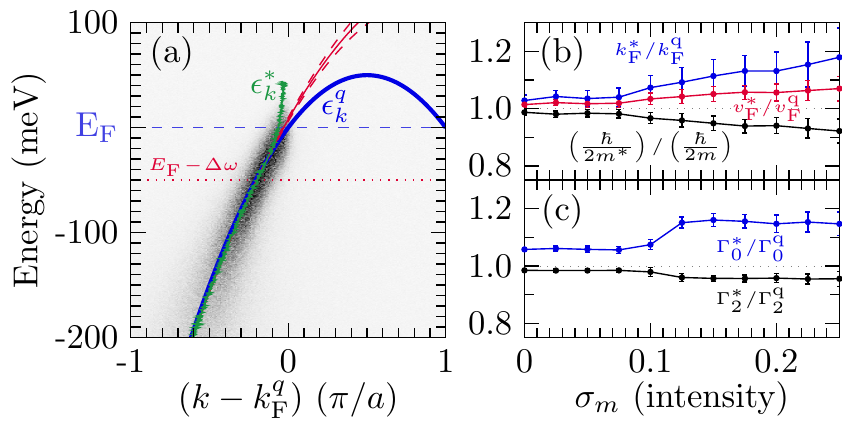}}
\caption{
\label{fig:snr} 
\figtitle{When a synthetic noise is included into the simulations, the quasiparticle dispersion extracted from the MDC analysis $\epsilon_k^*$, green line in (a), becomes jagged. A quadratic fit to $\epsilon_k^*$ defines a range of possible consistent dispersions which are shown by the red dashed lines.
The simulated ARPES spectra shown in a gray scale in (a) is based on the same parameters presented in Fig.\,\ref{fig:qbandqwidth}(a) with a noise level of 0.25 for the additive and multiplicative components. 
 The increase of the multiplicative noise level $\sigma_m$ affects the mean value of the obtained parameters from IDA, as shown by the data in (b). The error bars represent the standard deviation based on a statistical ensemble of 100 samples. The additive noise level is kept constant to $\sigma_a=0.25$.}
}
\end{figure}

Here we add synthetic noise to the simulations to study its effect on the IDA, and show that it leads to a quasi-linear increase in the standard deviation with noise level. We use an effective noise model with only additive and multiplicative components.\citep{ccd_noise}  
 The components are represented by two uncorrelated, stochastic, and normally-distributed variables with variance $\sigma$,  $N(\sigma_a)$ and  $N(\sigma_m)$, which are added to the signal $S$ obtaining a noisy signal $M$:
  \begin{equation}
  		M = S + S \times N(\sigma_m) + N(\sigma_a).
  \end{equation}
 We also assume that they are uncorrelated in energy and momentum. 
 The amplitude of the noise in this model is characterized by the width of the probability distribution which, in the case of a normal distribution, is given by the variance. Note that the values of the variance are reported normalized to the total intensity of the MDCs.
When we apply the noise model to the parabolic dispersion and quadratic linewidth shown in Fig. \ref{fig:qbandqwidth}(a), the extracted dispersion $\epsilon_k^*$,  shown in Fig.\,\ref{fig:snr}(a), becomes jagged. 
The set of curves produced by considering the error bars obtained from a quadratic fit to 
$\epsilon_k^*$ are bounded by the red dashed lines in Fig.\,\ref{fig:snr}(a). A fit to a noisy $\epsilon_k^*$ results in a standard error for the fit parameters and, as expected, this error increases linearly with $\sigma_m$. 

Because the IDA is a numerical transformation, we study a statistical ensemble of 100 samples to understand how the average and standard deviation of the parameters evolve with the noise level. 
The accuracy of the method is reflected in the evolution of the average which represents a systematic error, while the precision is determined by the standard deviation. We focus on their evolution with increasing multiplicative noise level $\sigma_m$ because it produces a bigger effect than the additive.  For an energy resolution of $\Delta \omega\!=\!50$\,meV and an additive noise level of $\sigma_a\!=\!0.25$, the average and standard deviation increase almost linearly with $\sigma_m$ as shown in Fig. \ref{fig:snr}(b), at least up to  $\sigma_m\!=\!0.25$. 
This exemplifies the effect of the IDA and shows that for the realistic parameters used, an accuracy of about $10\%$ is obtained for a noise level of $\sigma_a=0.25$ and $\sigma_m<0.1$, with a similar precision as defined by three times the standard deviation.

\bibliography{Levy_resolution_2014_biblio_v05}

\begin{thebibliography}{33}
\bibitem [{mom()}]{momres}%
  \BibitemOpen
  \href@noop {} {}\bibinfo {note} {In ARPES, there are two contributions to the
  momentum resolution: the angular resolution of the detector and the energy
  resolution of the monochromator. The relative momentum resolution for
  momentum parallel to the surface can be expressed as: \[ \frac{\Delta
  k_\parallel}{k_\parallel}=\sqrt{\left[\Delta \theta \cot(\theta)
  \right]^2+{R}^{-2}},\] where $\theta$ is the polar emission angle defined by
  the geometry of the experimental setup and ${R}$ is the monochromator
  resolving power. The first term, the angular contribution, dominates due to
  the high resolving power normally achieved, $R_m\!>\!100$.}\BibitemShut
  {Stop}%
\end{thebibliography}

\end{document}